\documentclass[preprint,preprintnumbers,amsmath,amssymb]{revtex4-2}

\usepackage{amsfonts}    
\usepackage{amssymb}
\usepackage{scrextend}
\usepackage{amsmath}
\usepackage{latexsym}
\usepackage{eepic}
\usepackage{graphicx}
\usepackage{color}

\usepackage[active]{srcltx}

\newcommand{\al}{\alpha}
\newcommand{\pa}{\partial}

\newcommand{\si}{\sigma}

\newcommand{\De}{\Delta}

\newcommand{\tha}{\theta}

\newcommand{\rar}{\rightarrow}

\def\be {\begin {equation}}
\def\ee {\end {equation}}

\newcommand{\ba}{\begin{array}}
\newcommand{\ea}{\end{array}}
\newcommand{\bea}{\begin{eqnarray}}
\newcommand{\eea}{\end{eqnarray}}
\newcommand{\bi}{\begin{itemize}}
\newcommand{\ei}{\end{itemize}}

\begin{document}

\title{Two-body Coulomb problem and $g^{(2)}$ algebra (once again about the Hydrogen atom)}

\author{Alexander~V.~Turbiner}
\email{turbiner@nucleares.unam.mx}
\affiliation{Instituto de Ciencias Nucleares, Universidad Nacional
Aut\'onoma de M\'exico, Apartado Postal 70-543, 04510 M\'exico,
D.F., Mexico}

\author{Adrian M. Escobar Ruiz}
\email{admau@xanum.uam.mx}
\affiliation{Departamento de F\'isica, Universidad Aut\'onoma Metropolitana-Iztapalapa,
Apartado Postal 55-534, 09340 M\'exico, D.F., Mexico}

\begin{abstract}

Taking the Hydrogen atom as an example it is shown that if the symmetry of a three-dimensional system is $O(2) \oplus Z_2$, the variables $(r, \rho, \varphi)$
allow a separation of the variable $\varphi$, and the eigenfunctions define a new family
of orthogonal polynomials in two variables, $(r, \rho^2)$. These polynomials
are related to the finite-dimensional representations of the algebra
$gl(2) \ltimes {\it R}^3 \in g^{(2)}$ (discovered by S Lie around 1880 which went almost unnoticed), which occurs as the hidden algebra of the $G_2$
rational integrable system of 3 bodies on the line with 2- and 3-body interactions
(the Wolfes model).
Namely, those polynomials occur intrinsically in the study of the Zeeman effect
on Hydrogen atom. It is shown that in the variables $(r, \rho, \varphi)$ in the quasi-exactly-solvable generalized Coulomb problem new polynomial eigenfunctions
in $(r, \rho^2)$-variables are found.

\end{abstract}

\keywords{}

\maketitle


\newpage

The Hydrogen atom is one of the most fundamental systems in quantum physics and in Nature.
Of course, it is present in practically all textbooks on quantum mechanics, see e.g. \cite{LL:1977}. Usually, exploring $O(3)$ geometrical symmetry, it is studied in spherical coordinates $(r, \tha, \varphi)$, see Fig.1, although parabolic coordinates $(r+z, r-z, \varphi)$ are also used in different applications. The states are characterized by the well-known Coulomb quantum numbers $(n_r, \ell, m)$ and the Laguerre polynomials $L_{n_r}(r)$ naturally occur in the eigenfunctions. In this Letter the coordinates $(r\,,\rho\,,\varphi)$ are used for which the eigenfunctions are characterized by new orthogonal polynomials in two variables $(r, \rho^2)$.

Let us take the two body Coulomb system or, equivalently, the Hydrogen-like system $(Ze,-e)$ in three-dimensional space.
After center-of-mass separation we arrive at the three-dimensional space of relative motion which will be parametrized by variables $(r\,,\rho\,,\varphi)$, see Fig.1 for geometrical setting, here $r=\sqrt{x^2+y^2+z^2}$ is radius, $\rho^2=x^2+y^2$, $\tan \varphi = y/x$, when expressed in the Cartesian coordinates of the space of relative motion.
\begin{center}
\begin{figure}[h]
\includegraphics[scale=1.0]{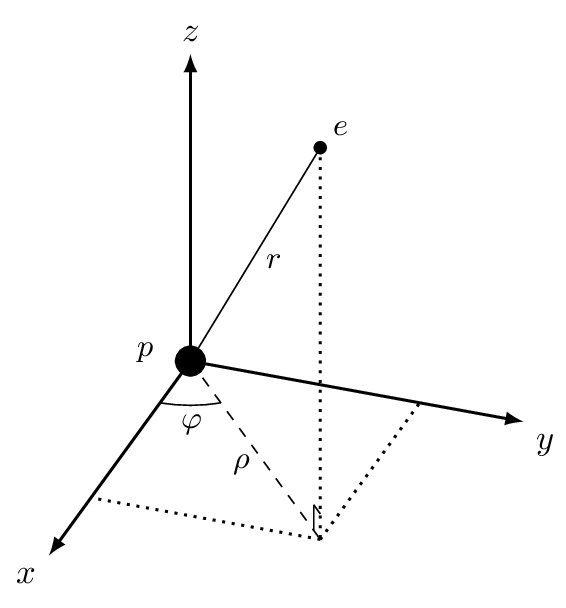}
\caption{Coordinates $(\rho,r,\varphi)$ at upper half-space $z \geq 0$. Center-of-Mass is located at the origin (marked by bullet and letter $p$).}
\label{fig:coordinates}
\end{figure}
\end{center}
The Hamiltonian in the relative space is of the form
\begin{equation}
\label{Hamiltonian}
{\cal H} \ = \ \frac{1}{2\mu}\,{{\bf \hat p}^2}\ - \ \frac{\al}{r} \ ,
\end{equation}
here $\alpha=Ze^2>0$ for the case of attractive charges, $\bf \hat p = -i \hbar \nabla$ is the momentum operator and we assume for simplicity a unit reduced mass, $\mu=1$ from now on and setting $\hbar=1$. Instead of solving the Schr\"odinger equation
\[
  {\cal H}\,\Psi\ =\ E\,\Psi\ ,
\]
we introduce the operator
\begin{equation}
\label{H}
  r\,({\cal H}\ -\ E)\ =\ \left(-\frac{r}{2} \De - E r\right)\ -\ \al \ \equiv (H - \al)\ ,
\end{equation}
and look for its zero modes, here $\De$ is the Laplacian. This is equivalent to the study of the spectrum of the operator $H$,
\begin{equation}
\label{H-equation}
H \,\Psi\ =\  \al \,\Psi\ ,
\end{equation}
where $\al$ - the Coulomb parameter - plays a role the spectral parameter, while the energy $E$ - the original spectral parameter - is considered fixed. This procedure resembles the introduction of the known Sturm representation in quantum mechanics.

Recently, in studying the Hydrogen atom in a constant uniform magnetic field
it was shown a convenience of use the coordinates $(r, \rho, \varphi)$ for the parametrization of the $SO(2) \oplus Z_2(z \rar -z)$-sub-symmetry of the system (\ref{Hamiltonian}), see \cite{dVTE:2021}. Such coordinates imply a partial separation of variables - the variable $\varphi$ can be separated out - and a representation of the wavefunctions of the Hydrogen atom as follows,
\begin{equation}
\label{wavefunction}
  \Psi\ =\ \rho^{|m|}\,e^{-\beta r} \,z^p\,e^{i\,m\,\varphi}\,P(r,\rho; m, p)\ \equiv \ \Gamma \,P(r,\rho; m, p)\ ,
\end{equation}
where $m$ is the magnetic quantum number, $z^2=r^2-\rho^2$, $p=0,1$ defines parity, and $\beta=\sqrt{-2E}$. Here, the first two factors emerge from the asymptotic behaviour of the solutions at small and large distances, respectively, whilst the factor $e^{i m \varphi}$ is the eigenfunction of the $z$-component of the angular momentum $L_z=-i \pa_{\varphi}$, which is conserved. Eventually, the problem of finding the eigenfunctions $\Psi$ is reduced to finding the eigenfunctions $P(r,\rho)$.

Let us take
\begin{equation}
\label{gf}
\Gamma \ = \ \rho^{|m|}\,e^{-\beta\,r} \,z^p\,e^{i\,m\,\varphi} \ ,
\end{equation}
as the gauge factor and introduce the gauge rotated operator
\begin{equation}
h \ \equiv \ \Gamma^{-1}\, H\,\Gamma  \ .
\end{equation}
After some calculations we arrive at
\begin{equation}
\label{ho}
\begin{aligned}
h \  = & \ -\,\frac{1}{2}\,r\,\pa^2_r \ - \ \frac{1}{2}\,r\,\pa^2_\rho \ - \ \rho\,\pa^2_{r,\rho}\ - \ \frac{(1+2 |m|)\,r\,-\,2\,\beta\,\rho ^2}{2\, \rho }\, \pa_\rho
\\ &
-( 1+p+|m| \,-\,\beta\,r )\,\pa_r \ + \ \beta \,(1+p+|m|)  \ .
\end{aligned}
\end{equation}
Since the operator is $Z_2$-invariant, $(\rho \rar -\rho)$, this suggests to introduce a new variable $u \equiv \rho^2$. In variables ($r,\,u$) the operator $h$ (\ref{ho}) takes the form of an algebraic operator with polynomial coefficients
\begin{equation}
\label{h-algebraic}
\begin{aligned}
h_a(r,u) \  = & \ -\,\frac{1}{2}\,r\,\pa^2_r \ - \ 2\,r\,u\,\pa^2_u \ - \ 2\,u\,\pa^2_{r,u}
\ - \ 2\,[(1+|m|)\,r\,-\,\beta\,u]\, \pa_u
\\ &
- \, ( 1+p+|m| \,-\,\beta\,r)\,\pa_r \ + \ \beta \,(1+p+|m|)\ ,
\end{aligned}
\end{equation}
and the eigenvalue problem (\ref{H-equation}) becomes
\begin{equation}
\label{h-equation}
h_a \,P\ =\  \al \,P\ ,
\end{equation}
where the domain is defined as $\{r \geq \rho \geq 0\}$, see Fig.2.
\begin{center}
\begin{figure}[h]
\includegraphics[scale=1.0]{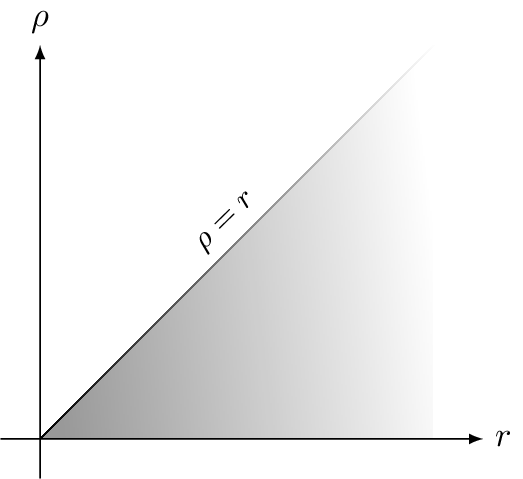}
\caption{Domain (shaded in gray) for the operator (\ref{ho}) in  $(r, \rho)$ variables, cf.\cite{dVTE:2021}.}
\label{fig:domain}
\end{figure}
\end{center}

It can be immediately checked that the operator $h_a$ has the infinitely-many invariant subspaces
\begin{equation}
\label{Pn}
   {\mathcal P}_n\ =\ <r^p u^q\ |\ 0 \leq p+2q \leq n >\ ,\ n=0,1,2,\ldots\ ,
\end{equation}
\[
   h_a:\,{\mathcal P}_n \ \rar {\mathcal P}_n \ ,
\]
see the Newton triangle on Fig.3 as illustration. Note that
the dimension of invariant subspace ${\mathcal P}_n$ (the number of points inside the Newton triangle including those on its boundaries),
\begin{equation}
\label{D}
    D\ \equiv \ \dim {\mathcal P}_n\ =\ \bigg[\frac{n}{2}\bigg]\bigg[\frac{n+1}{2}\bigg] \ + \ n \, + \, 1\ .
\end{equation}
This property implies that the eigenfunctions $P$ in (\ref{h-equation})
are polynomials, which allows us to calculate the eigenvalues in (\ref{h-equation}),
\[
   \al_n =\ \beta (n \ + \ 1+p+|m|)\ .
\]


\begin{center}
\begin{figure}[h]
\includegraphics[scale=0.5]{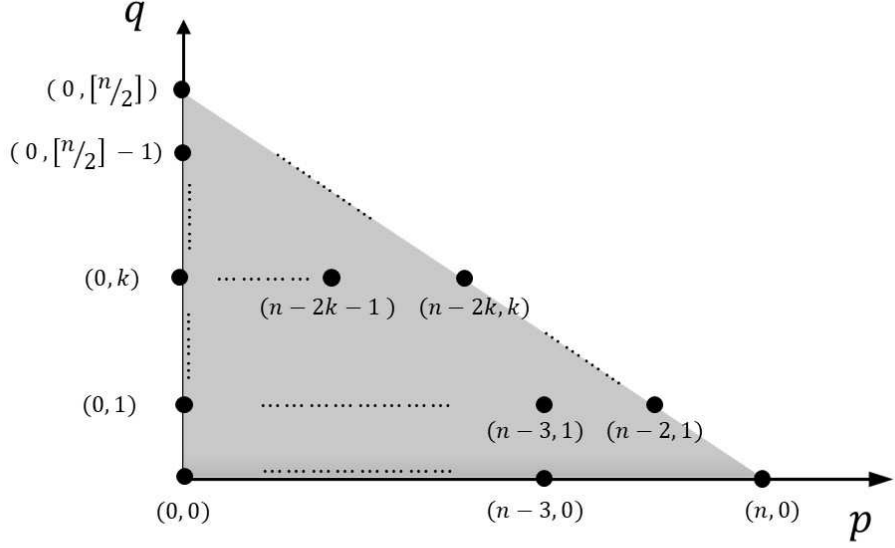}
\caption{The Newton diagram/triangle in lattice space where each monomial $(r^p\,u^q)$ is marked by point with coordinates $(p,q)$ illustrating the finite-dimensional space ${\mathcal P}_n$, which is an irreducible representation space of the algebras $g^{(2)}$ and a reducible one of the $gl_2 \ltimes R^3$.}
\label{fig:in1}
\end{figure}
\end{center}

\noindent
Assuming that $\al_n=\al$ is fixed, one can find the discrete values of the parameter $\beta$ in the operator (\ref{h-algebraic}) for which (\ref{h-equation}) holds. It defines the family of the operators $h_a$ with the same eigenvalue $\al$.  This leads to the well-known formula for energies,
\begin{equation}
\label{Ecoulomb}
     E\ =\ -\frac{\al^2}{2 (n\,+\,|m|+p+1)^2}\ ,
\end{equation}
where $n$ is the degree of the eigenpolynomial $P_n(r, u) \in {\mathcal P}_n$, see (\ref{Pn}). Hence, the eigenfunctions of the original Hamiltonian (\ref{Hamiltonian}) are now labelled by three quantum numbers ($n,p,m$), the combination $(n\,+\,|m|+p+1)$ plays the role of the principal quantum number, which is equal to $(n_r+\ell+1)$ in the usual Coulomb quantum numbers classification of states $(n_r, \ell, m)$.

{\it Examples I:}

\begin{itemize}
  \item $n=0$ \quad $\rar$\quad $\al_0 = \beta (|m|+p+1)\ ,$
\[
 E_0\ =\ -\frac{\al^2}{2 (|m|+p+1)^2}\ ,\ P_0\ =\ 1\ ,
\]
  \item $n=1$ \quad $\rar$\quad $\al_1 = \beta (|m|+p+2)\ ,$
\[
    E_1\ =\ -\frac{\al^2}{2 (|m|+p+2)^2}\ ,\ P_1\ =\ 1 - \frac{\al}{(|m|+p+1)(|m|+p+2)}\,r\ ,
\]
  \item $n=2$ \quad $\rar$\quad $\al_2 = \beta (|m|+p+2)\ ,$
\[
    E_2\ =\ -\frac{\al^2}{2 (|m|+p+3)^2}\ ,
\]
\[
\hskip -1cm
 P_2^{(1)}\ =\ 1 \  - \ \frac{2\, \alpha }{(|m|+p+1) (|m|+p+3)} \,r \ + \ \frac{2\, \alpha ^2}{(|m|+p+1) (|m|+p+3)^2 (2 |m|+2 p+3)}\, r^2 \ ,
\]
\[
\hskip -0.8cm
P_2^{(2)}\ =\ 1  \ - \ \frac{2\, \alpha }{(|m|+p+1) (|m|+p+3)} \,r \ + \ \frac{\alpha ^2}{(|m|+1) (|m|+p+1) (|m|+p+3)^2}\,u \ .
\]

This case $n=2$ indicates two-fold degeneracy.
\end{itemize}

In general, the polynomials $P_n$ are orthogonal with measure $\Gamma^2$, see (\ref{gf}). It seems evident that these polynomials form a complete basis, although the rigorous proof is absent. Those polynomials are not in the list of known orthogonal polynomials in two variables, see \cite{Krall:1938} and for a review \cite{Littlejohn:1988}. For a given $n > 0$, the degeneracy/multiplicity is equal to $n$ for even $n=2k$, while for odd $n=2k+1$ it is equal to $2k=n-1$: it is the number of points on the hypotenuse of the Newton triangle shown in Fig.3.

The invariant subspace ${\mathcal P}_n$ coincides with the finite-dimensional representation space of the algebra $g^{(2)}$: the infinite-dimensional, eleven-generated Lie algebra of differential operators introduced in \cite{Turbiner:1998} (see for a discussion \cite{Turbiner:2013t}) in relation to the $G_2$-integrable (both rational and trigonometric as well as elliptic) system of the Hamiltonian reduction. Generating elements are spanned by the Euler-Cartan generator
\begin{equation}
\label{J0-g2}
{\tilde {\cal J}}_0(n)\ =\ r\pa_r \  +\ 2 u\pa_u \ - \ n\ ,
\end{equation}
and
\[
 {\cal J}^1\  =\  \pa_r\ ,\
 {\cal J}^2_n\  =\ r \pa_r\ -\ \frac{n}{3} \ ,\
 {\cal J}^3_n\  =\ 2 u\pa_u\ -\ \frac{n}{3}\ ,
\]
\begin{equation}
\label{gl2r}
       {\cal J}^4_n\  =\ r^2 \pa_r \  +\ 2 r u \pa_u \ - \ n r\ =\ r {\tilde {\cal J}}_0(n)\ ,
\end{equation}
\[
 {\cal R}_{0}\  = \ \pa_u\ ,\ {\cal R}_{1}\  = \ r\pa_u\ ,\ {\cal R}_{2}\  = \ r^{2}\pa_u\ ,\
\]
\begin{equation}
\label{gl2t}
 {\cal T}_0\ =\ u\pa_{r}^2 \ ,\ {\cal T}_1\ =\ u\pa_{r} {\tilde {\cal J}}_0{(n)}\ ,\
 {\cal T}_2\ =\
  u{\tilde {\cal J}}_0{(n)}\ ({\tilde {\cal J}}_0{(n)} + 1) = \ u {\tilde {\cal J}}_0{(n)}\ {\tilde {\cal J}}_0{(n-1)}\ ,
\end{equation}
where $n$ is a parameter. If $n$ is a non-negative integer, the space ${\mathcal P}_n$ is the common invariant subspace for the generators (\ref{J0-g2}), (\ref{gl2r}), (\ref{gl2t}), where the algebra acts irreducibly.

The differential operator $h_a(r,u)$ can be written in the $g^{(2)}$-Lie-algebraic form,
\begin{equation}
\label{h-Lie-algebraic}
 h^{(g^{(2)})}_{a}\ =\ -\,\frac{1}{2}\,{\cal J}^2_n {\cal J}^1\ - \ {\cal J}^3_n\,
 {\cal R}_{1} \ - \ {\cal J}^3_n\,{\cal J}^1\ +\ \beta {\tilde {\cal J}}_0(n)\ -
\end{equation}
\[
   \left( 1+p+|m|+\frac{n}{2}\right)\,{\cal J}^1\ - \ 2\,\left(1+|m|+\frac{n}{6}\right)\,{\cal R}_{1}
   \ + \ \beta \,(1+p+|m| + n)\ ,
\]
in terms of the generators (\ref{J0-g2}), (\ref{gl2r}) only. Note that the generators (\ref{J0-g2}), (\ref{gl2r}) span the non-semi-simple Lie subalgebra $gl(2) \ltimes
{\it R}^3 \in g^{(2)}$, discovered by Sophus Lie circa 1880 \cite{Lie:1880} as the algebra of vector fields,  see \cite{gko:1992} (Cases 24, 28 at $r=2$), it was extended to first order differential operators in \cite{gko:1994}. {\color{blue} In general, the algebra (\ref{J0-g2}), (\ref{gl2r}) acts on the functions in two variables $(r,u)$. However, in the action on the functions in one variable $r$, this algebra acts as $sl(2)$-algebra,
\begin{equation}
\label{sl2}
  J^+_n = r^2\pa_r - n r\ ,\ J^0_n = 2r\pa_r - n\ ,\ J^- = \pa_r \ .
\end{equation}

The eigenpolynomials of the operator $h_a(r,u)$ (\ref{h-algebraic}) do not admit factorization $P(r,u) \neq Q_1(r) Q_2(u)$. However, it can be seen that in the family of polynomial eigenfunctions in two variables $(r,u)$, there exists an infinite (sub)-family of eigenpolynomials $P(r)$, which depends on the single variable $r$ {\it only}. In {\it Examples I} this subfamily is represented by $P_0, P_1, P_2^{(1)}$. In this case the operator $h_a(r,u)$ degenerates into the Laguerre operator
\begin{equation}
\label{har}
  h_a(r) \  = \ -\,\frac{1}{2}\,r\,\pa^2_r \
- \ ( 1+p+|m| \,-\,\beta\,r)\,\pa_r \ + \ \beta \,(1+p+|m|)\ ,
\end{equation}
and its eigenpolynomials are nothing but the associated Laguerre polynomials. It implies that $P_0, P_1, P^{(1)}_2$ are the first three (associated) Laguerre polynomials. In turn, the Laguerre operator has a Lie-algebraic form in terms of generators ${\cal J}^1, {\cal J}^2_0$, see (\ref{gl2r}),
\begin{equation}
\label{h-Lie-algebraic-Lag}
 h^{(g^{(2)})}_{a,Lag}\ =\ -\,\frac{1}{2}\,{\cal J}^2_0 {\cal J}^1\
 +\ \beta {\cal J}^2_0\ -\ \left( 1+p+|m|\right)\,{\cal J}^1\
   \ + \ \beta \,(1+p+|m|)\ .
\end{equation}
Let us note that ${\cal J}^1, {\cal J}^2_0$ span the Borel subalgebra $b_2$ of the $sl(2)$-algebra of the first order differential operators, $b_2 \in sl(2)$. Needless to say that in terms of the $sl(2)$ generators (\ref{sl2}), the operator (\ref{har}) has the form similar to (\ref{h-Lie-algebraic-Lag}).

The phenomenon of the existence of the eigenfunctions of the smaller number of variables than ones, which appear in the Hamiltonian, was systematically observed for many-body problems where the potential depends on relative distances alone \cite{twe}. Such a situation is typical for the case of separation of variables. But it is not at all typical when the variables are not separated. This phenomenon was called in \cite{twe} {\it beyond the separation of variables}.
}

{\color{blue}
\noindent
{\it Quasi-exact-solvability.}
It is known that the Coulomb potential admits quasi-exactly-solvable extension in radial direction \cite{QES:1988}. In order to see it let us take the operator $h_a$ (\ref{h-algebraic}) and make gauge rotation of it with Gaussian gauge factor,

\[
   \tilde h\ \equiv \ e^{A \frac{ r^2}{2}}\ (h_a + r \,W) \ e^{-A \frac{ r^2}{2}}\ = \ h_a (\pa_r \rar \pa_r - A\,r)\  + \ r\,W \ ,
\]
where $A>0$ and $W$ is a function to define. As a result we arrive at
\begin{equation}
\label{htilde}
  \tilde h\ =\ h_a(r,u)\ + \ A\,\bigg( r^2 \pa_r + 2  \,r\, u\, \pa_u - n \,r  \bigg) - \frac{A^2}{2} r^3
   + A\,r\, \bigg(\frac{3}{2} + n + p + |m| - \beta\, r\bigg) + r\,W \ .
\end{equation}
By choosing $W$ as follows
\begin{equation}
\label{W}
  W\ =\ \frac{A^2}{2} r^2\ - \ A\, \bigg(\frac{3}{2} + n + p + |m| - \beta \,r\bigg) \ ,
\end{equation}
allows us to vanish the 3rd and 4th terms in the r.h.s. of (\ref{htilde}) and we arrive at the quasi-exactly solvable operator in algebraic form (with polynomial coefficients)
\[
  \tilde h_n(r,u)\ =\ h_a\ + \ A\,\bigg( r^2 \pa_r + 2  r u \pa_u - n r  \bigg)\ =
\]
\begin{equation}
\label{hn}
-\,\frac{1}{2}\,r\,\pa^2_r \ - \ 2\,r\,u\,\pa^2_u \ - \ 2\,u\,\pa^2_{r,u}
\ -
\end{equation}
\[
 2\,[(1+|m|)\,r\,-\,\beta\,u - 2A r u]\, \pa_u
- \, ( 1+p+|m| \,-\,\beta\,r - A r^2)\,\pa_r \ + \ \beta \,(1+p+|m|) - A n r\ ,
\]
such that ${\mathcal P}_n$ (\ref{Pn}) is their invariant subspace
\[
   \tilde h_n:\,{\mathcal P}_n \ \rar {\mathcal P}_n\ .
\]
In the $g^{(2)}$-Lie-algebraic form the operator $\tilde h_n$ appears as follows,
\begin{equation}
\label{h-Lie-algebraic-qes}
 h^{(g^{(2)},qes)}_{n}\ =\ h^{(g^{(2)})}_{a}\ + A\  {\cal J}^4_n\ ,
\end{equation}
cf.(\ref{h-Lie-algebraic}), thus, it involves the ``positive-root", raising generator ${\cal J}^4_n$, see (\ref{gl2r}).
The corresponding spectral problem
\begin{equation}
\label{htilde-SP}
   \tilde h_n\, P\ =\ \al\, P\ ,
\end{equation}
where $\al$ is the spectral parameter, has a finite number of polynomial eigenfunctions in $(r,u)$ variables equal to the dimension of the invariant subspace $D$, see (\ref{D}).
It can be checked that among $D$ polynomial eigenfunctions in $(r,u)$ variables the spectral problem (\ref{htilde-SP}) has also $(n+1)$ polynomial eigenfunctions in the single variable $r$. Those polynomial eigenfunctions are already known as ones of the one-dimensional radial quasi-exactly-solvable problem \cite{QES:1988}. In this case the operator $\tilde h_n(r,u)$ (\ref{hn}) degenerates to
\[
  \tilde h_n(r)\ =\ -\,\frac{1}{2}\,r\,\pa^2_r\
- \, ( 1+p+|m| \,-\,\beta\,r - A r^2)\,\pa_r \ + \ \beta \,(1+p+|m|) - A n r\ ,
\]
cf. \cite{PhysReptsQES:2016}.
All other eigenfunctions of (\ref{htilde-SP}) are non-polynomials. It is worth presenting several polynomial eigenfunctions explicitly.

\bigskip

{\it Examples II}:
\begin{itemize}
  \item $n=0 \ (D=1)$
\[
    \al_0^{(qes)}\ =\ \beta \,(1+p+|m|)\qquad ;\qquad P_0\ =\ 1\ ,
\]
 \item $n=1 \ (D=2)$:\ (two eigenstates)
\[
    \al_{1,i}^{(qes)}\ =\ \frac{1}{2} \left(\beta+2\, \beta\,  \sigma _1 \pm\sqrt{4 \,A\, \sigma _1+\beta ^2}\right) \quad ;\quad P^{(i)}_1\ =\ \frac{\sigma _1}{ \beta -\al_{1,i}^{(qes)} } \ + \ r \ ,
\]
where $i=+,-$ and $$\si_k \equiv (k+|m|+p)\ .$$
 \item $n=2 \ (D=4)$:\ (four eigenstates)
\[
    \al_{2,i}^{(qes)}\ \equiv \ \al_{2,i} \ ,\quad i=1,2,3\ ,
\]
and $\al_{2,i}$ are the roots of cubic equation
\[
\al_{2} ^3\ - \ \al_{2}^2\,\beta \, \left( \sigma _1+ \sigma _2+ \sigma _3\right)\ + \ \al_{2}\,\big[ \beta ^2 \sigma _1 \sigma _2+\beta ^2 \sigma _1 \sigma _3+\beta ^2 \sigma _2 \sigma _3\,-\,A \left(4 \sigma _1+1\right)\big]
\]
\[
\ + \ A \,\beta\,  \sigma _1 \left(4 \,\sigma _1+5\right)\ -\ \beta ^3 \,\sigma _1 \,\sigma _2\, \sigma _3 \ = \ 0 \ ,
\]
accordingly
\[
     P_2^{(i)}\ =\ 2 \,A\, \sigma _1+A-\left(\al_{2,i} -\beta  \sigma _2\right) \left(\al_{2,i} -\beta  \sigma _3\right) \ + \  2 A \left(\al_{2,i} -\beta  \sigma _3\right)\,r \ - \ {2 A^2}\,r^2 \ ,
\]
here three eigenvalues $\al_{2,i}$ and three eigenpolynomials $P_2^{(i)}$ are related via analytic continuation in $A$.

\[
    \al_{2,4}^{(qes)}\ =\ \beta \,(3+p+|m|)\qquad ;\qquad  P_2^{(4)}\ =\ u\ - \ \frac{2 (|m|+1)}{2 |m|+2 p+3}\,r^2\ ,
\]
\end{itemize}
Note that the eigenpolynomials $P_0, P^{(\pm)}_1, P_2^{(1,2,3)}$ belong to the single-variable family, while the eigenpolynomial $P_2^{(4)}$ belongs to two-variable family.

Quasi-exactly-solvable extension of the operator $H$ (\ref{H}) can be easily constructed
\[
   {\hat H}_n\ =\ H + r\,W\ =\ H + \frac{A^2}{2} r^3\ - \ A\,r\, \bigg(\frac{3}{2} + n + p + |m| - \beta \,r\bigg),
\]
see (\ref{W}), which can immediately be converted into the quasi-exactly-solvable Hamiltonian
\begin{equation}
\label{Hqes}
  {\cal H}^{(qes)}(\al) \ \equiv \ \frac{1}{2}\,{{\bf \hat p}^2}\ - \ \frac{\alpha}{r} + W \ = \
  \frac{1}{2}\,{{\bf \hat p}^2}\ - \ \frac{\alpha}{r}\ - \ A \,\bigg(\frac{3}{2} + n + p + |m| - \beta\, r\bigg)\ +\ \frac{A^2}{2}\, r^2\ ,
\end{equation}
here the potential contains the $\frac{1}{r},\,r,\,r^2$ terms. This Hamiltonian is already known in literature as the quasi-exactly-solvable extension of the Coulomb problem of the second kind under the name {\it generalized Coulomb problem} \cite{QES:1988}, see for review \cite{PhysReptsQES:2016}, Case VIII, in the radial form when the (generalized) centrifugal barrier is explicitly included. The case (\ref{Hqes}) corresponds to the absence of the centrifugal term and the kinetic energy given by the standard three-dimensional Laplacian.
The Hamiltonian (\ref{Hqes}) depends on the parameter $\al$, which for fixed integer $n$ and parameter $\beta$ (hence, the energy, $E=-\beta^2/2$), is the eigenvalue in the spectral problem (\ref{htilde-SP}).
In general, the eigenvalues in the spectral problem (\ref{htilde-SP}) are different.
Different $\al$'s define different Hamiltonians.

The eigenvalue problem
\begin{equation}
\label{Hqes2}
  {\cal H}^{(qes)}(\al)\,\Psi\ =\ E\,\Psi\ ,\ E \ = \ -\frac{\beta^2}{2} \ ,
\end{equation}
admits the solutions in the form
\begin{equation}
\label{Psi-qes}
\Psi \ = \ \rho^{|m|}\,e^{-\beta \,r-A \frac{ r^2}{2}} \,z^p\,e^{i\,m\,\varphi}\,P(r,u) \ ,
\end{equation}
where the first $D$ eigenpolynomials are polynomials in $(r,u)$ variables
with the {\it same} energy $E=-\frac{\beta^2}{2}$. Hence, we have $D$ different Hamiltonians where in each of them we know one eigenstate explicitly in the form (\ref{Psi-qes}). It implies that new polynomial two-variable eigenfunctions $P(r,u)$ of the Hamiltonian (\ref{Hqes}) are found(!) in addition to already known one-variable ones $P(r)$.
}

As a conclusion we have to emphasize that in the case of potentials $V(r,\rho)$, which are invariant with respect to $O(2) \oplus Z_2$,  the representation (\ref{wavefunction}) can be generalized to
\begin{equation}
\label{wavefunction-gen}
  \Psi\ =\ \rho^{|m|}\,e^{-\Phi(r,\rho)} \,z^p\,e^{i\,m\,\varphi}\,F(r,\rho; m, p)\ .
\end{equation}
where $\Phi, F$ are function to find. This representation was successfully applied to the two-body neutral Coulomb system in a constant uniform magnetic field $\gamma$ for the case of two lowest energy states of positive/negative parity: $(n=0, m=0, p=0/1)$ \cite{dVTE:2021}, where $F=1$. By using the representation (\ref{wavefunction-gen}) the perturbation theory was easily developed,
\[
   \Phi(r,\rho)\ =\ \beta r + \gamma^2 \Phi_1(r,\rho) + \ldots + \gamma^{2k} \Phi_k(r,\rho)  + \ldots \quad ,
   \quad F(r,\rho)\ =\ 1 \ ,
\]
algebraically (for weak magnetic fields) up to a very high order and then few-parametric compact trial functions were built leading to highly accurate results for both energies and quadrupole moment for {\it arbitrary} magnetic fields. In the case of excited states with $n>0$ the prefactor $F(r,\rho)$ carries the information about the nodal lines in $(r,\rho)$-space. This prefactor can be found in the perturbation theory
\[
F(r,\rho)\ =\ P_n(r, u; m, p) + \gamma^2 P^{(1)}(r,\rho) + \ldots + \gamma^{2k} P^{(k)}(r,\rho)  + \ldots \ ,
\]
while both corrections $\Phi_k(r,\rho)$ and $P^{(k)}(r,\rho)$ are polynomials, their coefficients can be calculated by algebraic means. Usually, a polynomial correction $P^{(k)}(r,\rho)$ can be written as a finite linear superposition of the polynomials
$P_n(r, u; m, p)$. It will be done elsewhere.

If in (\ref{wavefunction-gen}) the exponential is chosen as
$\Phi(r,\rho) = \beta r + A r^2/2$ the quasi-exactly-solvable problem - the generalized Coulomb problem occurs with pre-factors in the form of the triangular polynomials $\{Q_n(r,u)\}_{i=1,\ldots D} \in {\mathcal P}_n$. In this case each $Q_n$ is a linear superposition of the eigenpolynomials $P_{0,1,2,\ldots n}$ of the original Coulomb problem.

\section*{Acknowledgments}

A.V.T. is thankful to P.~Iliev (GeorgiaTech, Atlanta, USA) and P.~Olver (University of Minnesota, USA) for mail correspondence and helpful discussions. This research was supported in part by the CONACyT grant A1-S-17364 and PAPIIT grant {\bf IN113022} (Mexico).

\end{document}